
\magnification=1200
\baselineskip=14pt

 1

 1

\font\Bold=cmbx10 scaled\magstep 1
\font\chapter=cmbx10 scaled\magstep 2
 5
 3
\def\impc{\,h{\,\rm Mpc}^{-1}}
\def\mpc{\,h^{-1}{\rm Mpc}}
\def\kms{\,{\rm km s^{-1}}}
\def\ocdm{\Omega_{\rm CDM}}
\def\ohdm{\Omega_{\rm HDM}}
\def\obaryon{\Omega_{\rm baryon}}

\def\w{\omega (\theta)}

\def\ra{\rangle}
\def\la{\langle}

\def\mpc{\,h^{-1}{\rm Mpc}}

\def\w{\omega (\theta)}

%
\def\ref{\parskip=0pt\par\noindent\hangindent\parindent
    \parskip =2ex plus .5ex minus .1ex}
\def\gs{\mathrel{\raise1.16pt\hbox{$>$}\kern-7.0pt
\lower3.06pt\hbox{{$\scriptstyle \sim$}}}}
\def\ls{\mathrel{\raise1.16pt\hbox{$<$}\kern-7.0pt
\lower3.06pt\hbox{{$\scriptstyle \sim$}}}}
\topskip1.5cm
\centerline{\chapter
The large-scale structure in a universe}
\bigskip
\centerline {\chapter dominated by cold plus hot dark matter}
\bigskip
\bigskip
\centerline {\bf Y.P. Jing$^{1,2,3}$, ~~H.J. Mo$^{4,3}$,
{}~~G. B\"orner$^{3}$ \& ~~L.Z. Fang$^{2,5}$}
\vskip 2.5cm
\item{1} SISSA--International School for Advanced Studies,
Via Beirut 2, I-34014 Trieste, Italy
\item{2} Department of Physics, University of
Arizona, Tucson, 85721, USA
\item{3} Max-Planck-Institut f\"ur Astrophysik,
Karl-Schwarzschild-Strasse 1,
W-8046 Garching bei M\"unchen, Germany
\item{4} Institute of Astronomy,
Madingley Road, Cambridge CB3 0HA, England
\item{5} Steward Observatory, University of
Arizona, Tucson, 85721, USA
\vskip 1.5cm
\centerline{Accepted for publication in: {\it Astronomy and Astrophysics}}
\bigskip
\noindent {{\bf Thesaurus codes}: 03.04.1 --03.12.1 --03.12.2 --07.08.1
--20.01.2}
\vskip 1cm
\centerline{\it All correspondence to: Y.P. Jing in USA}
\vfill
\eject
\baselineskip=14pt

\centerline{\Bold Abstract}
Using numerical simulations,
we investigate the large-scale gravitational clustering in
a flat universe dominated by cold plus hot dark
matter (i.e., $\Omega_0=\ocdm+\ohdm+\obaryon=1$).
Primordial density fluctuation spectrum is taken
to have the Zel'dovich-Harrison form.
Three models are studied, with
Model I having
$\ocdm=0.69$, $\obaryon=0.01$, and $\ohdm=0.30$ in one
flavor of neutrinos;
Model II having
$\ocdm=0.60$, $\obaryon=0.10$, and $\ohdm=0.30$ in one
flavor of neutrinos;
Model III having
$\ocdm=0.69$, $\obaryon=0.01$, and $\ohdm=0.30$ in three
flavors of neutrinos. The initial density spectra are
normalized by the COBE quadrupole measurement, and galaxies
are identified from the peaks of initial density fields above a certain
threshold chosen, to match the observed two-point correlation on
scales $\ls 10\mpc$. Thus the clustering properties of both the
mass and the galaxies are completely specified. The biasing
parameter (for the `galaxies')
determined in this way
is $b_g\approx 1.2$ for Model I, 1.5 for Model II
and 1.6 for Model III.

The clustering and motions of the simulated `galaxies' are
compared with recent observations.
The spatial distributions of galaxies in the hybrid models
are very frothy; filaments, sheets, voids etc. of sizes
10 -- 50$\mpc$ are frequently seen in the simulations.
All three models are in good agreement with the observed
local bulk motions and with the count-in-cell statistics $\sigma^2(l)$
in redshift surveys of IRAS galaxies. One exception is
the $\sigma^2(l)$ of the QDOT survey at $l=40\mpc$: the value is too
high to expect in the models. But its statistical significance
was recently questioned with an analysis of the 1.2 Jy IRAS survey.
Model I does not have
sufficient large-scale power to explain
the two-point angular correlation function of the APM
survey, the two-point correlation function of Abell clusters. Furthermore,
its galaxy pairwise velocity dispersion around $1\mpc$ is too high to
reconcile with the observation. The other two models can be
adjusted, within the observational errors, to fit all observations
on scales from $\sim 1\mpc$ to $\sim 50\mpc$, showing that the
power spectrum of the initial density fluctuation
is close to the predictions of these
two models, and indicating that the observational results of
galaxy clustering and motions on large scales are
consistent under some reasonable theoretical assumptions.

\vfill
\eject
\leftline{\Bold 1. Introduction}
\bigskip
Hybrid models, in which the universe is dominated by
cold plus hot dark matter (CDM and HDM, respectively),
were proposed in the
early 1980s as one scenario of
structure formation in the universe
(Fang et al. 1984; Shafi \& Stecker 1984;
Valdarnini \& Bonometto 1985).
These models were not considered very seriously,
however, because they depend on
at least one more parameter (i.e., the relative fraction of HDM and CDM)
than the simplest models
in which the universe is dominated by
a single kind of dark matter (e.g., CDM or HDM),
and there seemed then to be
no appealing reasons for studying them
in much more detail. The situation has changed recently,
because observational evidence has been accumulated
to the point where
one can conclude that the simplest CDM and HDM models do not
work.
The hybrid models are one of the simplest revisions of these
models.

The standard CDM model (e.g., Davis et al. 1985)
had been quite successful in explaining
the structure of the universe on scales $\ls 10\mpc$
(see Davis \& Efstathiou 1988 for a review).
The success was, however, based on the assumption
that galaxies are highly biased tracers
(with a bias parameter $b_g\sim 2.5$)
of the underlying mass distribution.
Since a strong
segregation in the clustering amplitude between
faint and bright galaxies is not observed
(Alimi et al. 1988; Mo \& Lahav 1993),
a high value of the bias is therefore not
favored. Furthermore, the high-bias model
fails to provide sufficient power on large
scales to match the angular correlation
functions of deep
galaxy surveys (Maddox et al. 1990,
hereafter MESL;
Picard 1991; Collins, Nichol, \& Lumsden 1992), the
large-scale velocity fields (e.g., Lynden-Bell et al.
1988; Bertschinger et al 1990), the
correlation functions of clusters
of galaxies (e.g., Bahcall \& Soneira 1983; Klypin \& Kopylov
1983; Postman et al. 1986; Batuski et al. 1989; Huchra et al. 1990;
Postman et al. 1992; Mo et al. 1993; Jing \&
Valdarnini 1993; Dalton et al. 1992; Nichol et al. 1992), and the COBE
measurement of the microwave background
radiation (MBR) fluctuations (Smoot et al. 1992).
In fact, the COBE result
is very close
to the prediction of the standard CDM model
with $b_g\approx 1$.
But the unbiased CDM model has serious problems
on small scales; the
predicted amplitude of the velocity field on Mpc scales is too
large to be
compatible with observations,
unless a large velocity bias exists between
galaxies
and dark matter
(Couchman \& Carlberg 1992).

The problems in pure HDM models are,
to some extent, the opposite. The free-streaming
motions of neutrinos (of mass $m_\nu$) have
erased all fluctuations on
scales smaller than $\sim 40(30{\rm eV}/m_\nu)\mpc$,
and galaxy formation has to
invoke the fragmentations of large pancakes.
To have galaxies (quasars) form early enough
to be consistent with observations, the amplitude
of the primordial density spectrum on large
scales would be too
large to be compatible with the upper limit of the MBR
anisotropy  on angles of $\sim 1^\circ$ (Bond et al. 1991).
Furthermore, the potential wells provided by neutrino
pancakes are so deep that
baryons
falling into clusters of galaxies
would exceed the observational
limits of the x-ray background (e.g.
White et al. 1984).

If the formation of structure is mainly due to
gravitational instability, one of the simplest
cases to study next would be hybrid models,
in which the universe is dominated
by CDM plus HDM.
{}From the above discussion we can imagine that the
difficulties in one model (CDM or HDM) could,
to some extent, be overcome in the other, and a
hybrid model may do a better job in matching
observations than the
simplest models do. Indeed, calculations
of the linear evolution of the density
perturbations have shown that the hybrid models,
when normalized on large scales, have less small-scale power
than a pure CDM model but much more than a
pure HDM model (e.g. Holtzman 1989; Xiang \& Kiang 1992;
van Dalen \& Schaefer 1992; Taylor \& Rowan-Robinson 1992),
which is in the desirable direction.
Davis et al. (1992) and Gelb et al. (1993) recently have performed
N-body simulations, with emphasis on the pairwise velocity in
the hybrid models.
Compared with the observed large-scale structure,
the results of all these
investigations suggest that the favorable
(hybrid) models
would have $\ohdm \sim 0.3$.

In order to study these models further, we have carried
out numerical simulations for three hybrid models
with $\Omega_0=1$ and
$\ohdm =0.3$. The three models differ in the relative
fraction of baryons and in the number of flavors of massive
neutrinos. Since our simulations do not treat
CDM and HDM separately, the results depend
only on the initial density spectra on scales larger than
$\sim 0.5\mpc$ (the resolution limit of the simulation of
box size $60\mpc$),
and they are not expected to
be valid on galactic scales. The amplitudes of the initial density
spectra are normalized by the COBE measurement of the fluctuation
in the MBR (Smoot et al. 1992).
Galaxies are identified from the density peaks above
certain threshold, to match the observed two-point correlation
functions on scales $\ls 10\mpc$. The large-scale clustering
and motions of `galaxies' are then determined in the models,
which enables us to make direct comparisons with observations.

The outline of the paper is as follows. In \S 2, we
describe in detail the models to be studied and the
simulations to be used to trace the structure evolution.
The identification procedure of galaxies is also
described there. In \S 3, we present the results of
our simulations, and compare them with various
observations of galaxy clustering and motions.
The clustering properties of rich clusters in the simulations
have been analyzed elsewhere (Jing et al. 1993; hereafter JMBF).
In \S 4, we give a brief discussion of our results
and summarize our main conclusions.

\bigskip
\bigskip
\leftline{\Bold 2. Models and N-body simulation}
\bigskip
\leftline{\bf 2.1 Models and simulation method}
\smallskip
In this paper, we assume that the universe
is flat (i.e. the total mass density
parameter $\Omega _0=1$), the cosmological
constant $\Lambda =0$, and the Hubble constant
$H_0=50\kms {\rm Mpc}^{-1}$ (written as
$H_0=100 h \kms {\rm Mpc}^{-1}$). The primordial
density fluctuations are assumed to be adiabatic and
Gaussian, with a Zel'dovich-Harrison spectrum $P(k)\propto k$.
The universe is assumed to be consist of
CDM,
three species of neutrinos
(the massive part of which is called HDM), and baryons.
In this framework, we study three models with
different relative fractions of HDM, CDM and baryons.
In Model I, the cosmic mass has $69\%$ in CDM,
30\% in one flavor of neutrinos and 1\% in baryons.
In Model II, 60\% of the cosmic mass is in CDM,
30 \% in one flavor of neutrinos and 10\% in
baryons. Model III has the same mass composition as
Model I, except that it contains
three flavors of neutrinos with equal mass.
The choices for the relative fractions of CDM and HDM
are based on the results of previous studies
of the hybrid (CDM$+$HDM) models
(e.g. Holtzman 1989; Xiang \& Kiang 1992;
van Dalen \& Schaefer 1992;
Wright et al. 1992; Davis et al. 1992;
Efstathiou et al. 1992; Taylor \& Rowan-Robinson 1992;
Gelb et al. 1993). By comparing models with the
observed large-scale structure, these studies suggested
that the (most) favored hybrid models would have
$\ohdm \sim 0.3$. The inclusion of a baryon component
generally suppresses the power on small scales.
The two values
$\obaryon=0.01$, 0.1 represent the lower and upper limits
given by the standard Big Bang
nucleosynthesis calculation (e.g. Olive et al. 1990;
Walker et al. 1991).
In Model I and Model II, the mass of the neutrinos is about
7eV, which is well below the experimental upper limit
on the mass of $\nu_\tau$
(see e.g. Primack 1992 for a recent summary
of the experimental limits on neutrinos masses).
The neutrino mass in Model III is about 2eV in each
flavor. Since the constraints on the masses of $\nu_\mu$ and
$\nu_e$ are quite stringent, this model may not
be favoured by particle physics.
However, since
the neutrino mass in this model is smaller, it has
less power on small scales.

We will use the transfer functions of the linear
density perturbations given by Holtzman (1989)
for these models.
Holtzman fitted
the transfer functions at the present time ($z=0$)
by a parameter-fitting function.
For wavenumbers $k\le k_{max}=1.24{\rm Mpc}^{-1}h$ ($h=0.5$),
the fitting function was claimed to have accuracy better than
ten percent (i.e. the maximum deviations of the $P(k)$
fit are less than 20\%, see Holtzman 1989).
As we shall see below,
the Nyquist wavenumber in our main set of simulations
(in a cube of $240\mpc$ in each side) is about
$k_N\approx 0.8 h {\rm Mpc}^{-1}$, well below the value of
$k_{max}$. For the other set where the cube
size is $60\mpc$, $k_N\approx 3 h{\rm Mpc}^{-1}$.
In this case, we simply
extrapolate the fitting functions to $1.24\le k\le 3.0 \impc$.
van Dalen \& Schaefer (1992)  and Klypin et al. (1993) have
given the transfer
functions for Model I and Model II up to $k=3\impc$, which
confirmed that
the extrapolations are correct within the ten percent
accuracy.
The linear density power spectra
for these models, given by
the parameter fitting, are shown in Figure 1.

Our simulations are performed by a particle-mesh code with $128^3$
grid points and $64^3$ particles in cubic comoving volumes (Hockney \&
Eastwood 1981). The standard Cloud-In-Cell (CIC) scheme is used for
mass assignment and force interpolation.
The Poisson equation for the gravitational potential
is solved by the Fast Fourier Transformation and the gravitational force
is calculated from the potential by the staggered-mesh method
(Melott 1986). Particle velocities and positions
are forwarded by the standard leapfrog integration. The
integration variable is the scale factor $a$, and the integration
step size $\Delta a$ is $0.1a_i$ ($a_i$ is the scale factor
at the initial time). The simulations are started at $z=8$, so there are
80 steps for the systems to evolve up to the present time $z=0$. Initial
velocities and positions of particles are generated by the
Zel'dovich approximation, following the prescription of Efstathiou
et al. (1985). The power spectra shown in Figure 1 are used to
produce the initial conditions of our simulations. The $P(k)$
are so normalized that the Sachs-Wolfe effect produces the
quadrupole $Q=6\times 10^{-6}$ detected by the COBE. Because
the simulations start at redshift $z=8$, we simply scale the
linear $P(k)$ by $P(k,z)\propto (1+z)^{-2}$ to get $P(k)$ at
$z=8$ (see discussions of \S 2.2).
For each model, we have done two sets of
simulations, using box sizes $L=60\mpc$ and $L=240\mpc$ respectively.
We run five realizations for each simulation.
For convenience,
we shall call the simulations of
the larger $L$ the L240 simulations, and those of the smaller $L$
the L60 simulations.
The mass of each particle is about $3\times 10^{13} {\rm M}_\odot$
in the L240 simulations and $5\times 10^{11} {\rm M}_\odot$ in
the L60 simulations.
The typical spatial resolution of PM codes is one mesh size, so
the L240 simulations have a resolution of
about $1.9\mpc$,  and the L60 simulations
about $0.5\mpc$.
Combining the results of
these two sets of simulations, we can obtain
sufficient resolution for the purpose of this paper,
the study of clustering features on scales
$0.5\ls r\ls 50 \mpc$.

\bigskip
\leftline{\bf 2.2. The validity of the simulations}
\smallskip
The linear power spectra  given
by Holtzman (1989) are
for redshift $z=0$.
The power spectra
at $z=8$ (when we start our simulations)
differ from
the simple scaling
$P(k,z)\propto (1+z)^{-2}$ for $k>k_J$, where $k_J$ is
the (comoving) Jeans wavenumber of neutrinos.
At
$z=8$, the
Jeans wavenumber is
$1.9 h {\rm Mpc}^{-1}$ for $m_\nu =7$eV, and
$0.5 h {\rm Mpc}^{-1}$ for $m_\nu =2$eV.
Since the Nyquist wavenumbers of the L60 and L240 simulations
are 3 and 0.8 $\impc$ respectively, the $(1+z)^{-2}$ scaling
suffices for the L240 simulations of Model I and Model II.
For the L60 simulations,
the use of the above
scaling will underestimate the power of clustering at $z=8$  near
the Nyquist wavenumber $k_N$ because of the neutrinos free-streaming
motion between $z=8$ and $z=0$.
The influence of the underestimation on our final results was tested
when we got the linear power spectra $P(k)$ of Model II at
different $z$ (Klypin et al. 1993, hereafter K93) after submitting the
first version of the paper. We did two
simulations of box size 60 $\mpc$. The method for
doing these simulations is the same as described above.
All input parameters (including
the random phases in density fields) are kept the same for the two
simulations except the initial power spectra are different. In one
simulation (NS simulation) we use the spectrum at $z=8$ given by K93, and
in the other (S simulation) we use the spectrum obtained by the
$(1+z)^{-2}$ relation from the spectrum at $z=0$ given by K93.
So the method used for the
S simulation is exactly the same as that
for other simulations
presented in this paper. To compare the two simulations,
we show their power spectra at several redshifts in Figure 2a. The
method for estimating the power spectra will be given in \S 3.1.
For the resolved range of the simulations, the largest difference
between the two spectra happens near the Nyquist wavenumber $k= k_N$.
At the initial time $z=8$, the $P(k)$ of the S simulation is about
40\% lower than that of the NS simulation at $k_N$.
The difference, however, becomes smaller with the development
of non-linear clustering, and is only about 15\% at $k=k_N$
when $z=0$.
This is the case because the
evolution of large-wavelength fluctuations
contributes to the (non-linear) clustering on small scales.
As Little et al. (1991)
showed, the non-linear clustering on $k\gs k_c$ is
mainly determined by the power
on $k\ls k_c$, where $k_c$ is the transition scale
from linear to non-linear clustering and is defined as $k_c=2\pi/R_c$ ($R_c$
is the radius of sphere in which the $rms$ density
fluctuation is 1). In our simulations,
$R_c$ is about $5\mpc$.
The initial difference in $P(k)$ is less than 20\% for
$k\ls k_c$. This is why the final difference of $P(k)$ in the two
simulations is only 15\% or smaller.

Another problem is associated with the neutrino free-streaming
motion, because we follow the evolution of structures by
only one type of particles (i.e. cold particles).
This procedure should be a good approximation
for wavenumbers less than
$k_J$,
so it is valid for the L240 simulations
of Model I and Model II.
But it may fail for the L60 simulations. As a check,
we have run a two-component simulation with box size
$60\mpc$ for Model II.
Because of the free-streaming
motion of the hot component, the usual way to set
the initial conditions
for simulations by using the Zel'dovich approximation is
no longer valid.
In Davis et al. (1992),
the initial condition is generated by
simply spreading the hot particles uniformly in the simulation
box and giving each of them a randomly-oriented velocity
drawn from a Fermi-Dirac distribution. This is obviously
an approximation, though the approximation may not be too bad
because of their small simulation box ($7\mpc$ on each side).
In K93, the authors claimed that
the Zel'dovich approximation was used to set the initial
conditions for their simulations,
but did not give any details.
We believe that
the use of the Zel'dovich approximation in
this context is not justified.

To test the influence of
the hot component on our results of the L60
simulations, we adopt
an `approximate' method to generate the initial conditions for the
two-component simulations.
The method
is still based on the Zel'dovich approximation but
requires that 1.) the initial position displacement of cold particles
corresponds to a random realization of the power spectrum $P_c(k)$
of cold particles; 2.) the initial position
displacement of hot particles
corresponds to a random realization of the power spectrum $P_h(k)$
of hot particles; 3.) the initial velocities of both cold and
hot particles, contributed by gravitational clustering, are assigned
by the usual Zel'dovich approximation,
assuming a power spectrum
$P(k)=(0.3
\sqrt{P_h(k)}+0.7 \sqrt{P_c(k)})^2$ and
a flat universe dominated by
cold dark matter; 4.) in realizing the above three steps, the
random phases of both CDM and HDM perturbations are kept the
same, so that they correspond to
a single random process; and 5.) each hot particle is given a thermal
motion randomly oriented and drawn from the Fermi-Dirac
distribution. We believe that
all of the above requirements are correct except the
third one.  The third is incorrect in general sense because
of the free-streaming motion of neutrinos, but may be not
a bad approximation for the L60 simulation,
for a large fraction of the peculiar
velocity induced by gravitational instability
in the linear regime is from
long-wavelength fluctuations and the free-streaming motion
affects only short-wavelength fluctuations
on a few Mpc after $z=8$.
Although we are unclear how accurate
the initial conditions generated
in this way can be, the test presented here may
give some idea about the importance of
the free-streaming motion in the simulations.

In the simulation, we use three sets of $64^3$ particles:
$64^3$ cold particles and $2\times 64^3$ hot particles.
The mass of each cold particle is $3.5\times 10^{11}
M_\odot$ and each hot particle has mass of $7.5\times 10^{10}
M_\odot$. The initial $P(k)$ at $z=8$ of both cold and hot particles
are from K93.
The hot particles are grouped into pairs at the initial time.
The two particles of each pair have the same {\it initial} position,
the same {\it initial}
velocity induced by the gravitational clustering, but
oppositely directed thermal velocities of equal magnitude.
As suggested by
K93, this may prevent the simulated thermal motions from
generating spurious fluctuations.
The random phases of the density fluctuations are designed
to be the same as in the NS simulation, so that these
two sets of simulations can be compared directly.
The simulation is evolved to the present time by the way
described earlier.

In Figure 2b, we present the power spectra $P(k)$ for both
cold and hot particles at several redshifts. For
comparison, the power spectra of the NS simulation
are also plotted.
By design, the power spectra of the cold particles in both
simulations are identical initially. The
hot particles have much less power on small scales
owing to
the free-streaming motion.  Because of the hot
component, the clustering of cold particles
becomes subsequently weaker
in the two-component simulation than in the NS simulation,
as is clearly shown by the power spectra
at redshift $z\ge 1$.
At redshift $z=1$, the
$P_c(k)$ of the two-component
simulation is lower, near the Nyquist wavenumber, than
that of the NS simulation by 25\%, an
amount comparable to the result based on the
linear calculation.
At $z\simeq 1$, the
Jeans wavenumber of neutrinos becomes $4\impc$, similar to the
force resolution size of  the simulation.
The free-streaming motion
is no longer important later on,
and the hot component catches up with the clustering
of the cold component in the two-component simulation.
In the meantime,
non-linear clustering becomes more and more important.
As discussed in the last paragraph, long wavelength perturbation
can significantly influence the clustering at short wavelengths.
As a result, the clustering difference between
the distributions of the cold particles in the
two sets of simulations becomes smaller at $z<1$. At the end
of the simulations, the $P_c(k)$ in the two-component simulation
is only $\sim 5\%$ lower than that in the NS
simulation.

The power spectrum of hot particles, as shown in Figure 2b, appears
to grow much faster near $k_N$ than that of cold particles,
especially at large redshift $z$.
This unphysical behavior is of common nature of N-body simulations.
Because the simulations consist of limited number of particles, shot
noise must play role to some extent. At the beginning of simulation,
particles are usually placed regularly, thus minimizing the
white noise. If the particles are cold, when clustering is
weak (linear), particles are moving coherently and the shot noise
is still largely suppressed (because particles are not random in space);
with clustering increasing, the shot noise become even less important.
But in the simulation of hot particles,
the behavior appears much more significant for their thermal
motions.  Because they have a component of random motion, they
become more or less randomly distributed in space on scale of the
random motion.
We should point out that the power spectra shown in the figures are
superpositions of physical clustering and white noise.
This is why the power spectrum of neutrinos appears
to grow much faster than that of cold particles, especially at large z.
But this does not mean that neutrinos really cluster faster than
cold dark matter.

The shot noise effect has little influence on our final conclusions
made in the paper.
In fact, we did another simulation which has two times more hot particles.
We  found that the shot noise effect of hot particles
is much less significant at $z=4$ , 2, and 1, as expected. But anyway,  this
effect is still showing up. However, the most interesting point is that
the distributions of cold particles in the two simulations are identical
at all redshifts (difference in $P(k)$ never exceeds $1\%$) and that the
distributions of hot particles are nearly identical at the final
stage of the simulations (difference in $P(k)$ is only 2\% at $k=k_N$).
By the way, the fact that shot noise does not amplify clustering
in the N-body simulations was noticed by Efstathiou et al. (1985)
and Davis et al. (1985).

{}From the above tests, we expect that, in the L60 simulations of
Model I and Model II, our neglect of a hot component
overestimates $P_c(k)$
by $\sim 5\%$.
And since the scaling $P(k,z)\propto (1+z)^{-2}$ for getting $P(k)$
at $z=8$ underestimates
$P_c(k)$ by $\sim 15\%$, the net systematic
errors arising from the two effects amount to 10\% in the power spectrum
measurement. This error is smaller than the fitting error of $P(k)$
in Holtzman (1989). At the present,
we are unclear how large an error
these two
effects will lead to
in the simulations of Model III. The error should
be larger,
because of the smaller neutrino mass.
As pointed out before, this model is not favored by particle physics.
We will consider this model as a phenomenological model
described by the power spectrum used in this paper.
\bigskip
\leftline{\bf 2.3 A simple biasing prescription}
\smallskip
Since the amplitude of the
initial density power spectrum is
normalized by the
COBE quadrupole observation, the clustering strength of the
underlying mass at the present time is uniquely determined.
Using the two-point correlation function $\xi$ as a clustering
measure, we found that the mass particles in all
three models are less
clustered than the galaxies in the universe (see \S 3.2).
We therefore
introduce a biasing mechanism to select galaxies in our
simulation, so that the two-point
correlation functions $\xi(r)$
of the simulated galaxies
have amplitudes $r_0$  similar to those
of the observed galaxies (Groth \& Peebles 1977; Davis \& Peebles 1983).
It should be pointed out that
the biasing mechanisms in galaxy formation theories
are not well established. They may
depend on many complicated physical processes
(e.g., Dekel \& Rees 1987). To identify galaxies in our
simulation, we follow the simple but plausible prescription of
White et al. (1987, hereafter WFDE). The key idea of this
prescription is that only
peaks above a certain threshold $\nu_s$ of the density field
smoothed on galactic scales ($r_s$)
will eventually evolve into galaxies
observed today (e.g., Kaiser 1984;
Bardeen et al. 1986; the smoothed field will be called $F_s$ below).
Because the particle distribution is not resolvable
on the scale $r_s$ in the L240 simulations, we cannot
identify the density peaks (or `galaxies') directly.
Instead, we first smooth the
density field [given by $P(k)$] on a much larger scale $r_b$ ($\gg r_s$) to
produce a background field $F_b$ which
is able to be resolved by the simulation, and
then use the analytical formula of Bardeen et al. (1986) to calculate the
expected peak density $n(\ge \nu_s)$ of $F_s$ in the vicinity of a point
where the density contrast of $F_b$ is known.
In implementing this method, we obtain first the density contrasts of
$F_b$ on the $128^3$ grid points and then the expected number of
peaks on each grid point. Because particles are uniformly placed on centers of
meshes before being perturbed,
the expected peak number $N_p$ associated with each particle is
just the sum of the
expected peak numbers of eight neighbouring grid points.

In our simulations, we have used a Gaussian window of
width $r_s=0.54\mpc$ to produce the smoothed field $F_s$. The background
field $F_b$ is generated by smoothing the original density field using a
top-hat window of radius $k_b=0.72\impc$ in $k$-space [i.e., $W(k)=1$
for $k\le k_b$; $W(k)=0$ for $k> k_b$].
As emphasized in WFDE, using the top-hat window
ensures that the clustering
properties of the peaks identified
above are asymptotically the same as those
of the {\it real peaks} of $F_s$.
The parameters $r_s$ and
$k_b$ are the same as in WFDE. The $\nu_s$ values are adjusted
such that the correlation length $r_0$ of peaks (i.e. $
\xi(r_0)=1$)
is about $6.0 \mpc$ (Davis
\& Peebles 1983; de Lapparent, Geller \& Huchra 1989). The
$\nu_s$ values thus determined are $0.0$
for Model I, $0.6$ for Model II, and $0.8$ for Model III. The mean
number densities of peaks are $0.019 \,h^3{\rm Mpc}^{-3}$ (Model I),
$0.015 \,h^3{\rm Mpc}^{-3}$ (Model II), and $0.016 \,h^3{\rm Mpc}^{-3}$
(Model III), all comparable to the observed density of galaxies brighter
than $M_{b_j} \approx -17.7$ (e.g., Loveday et al. 1992).
Our choices for the amount of biasing (i.e. $\nu_s$) are only for fitting
the observed correlation function of galaxies. No attempt has been made to
find any physical argument to support these choices,  because
physics on possible biasing processes is far unclear. The mean densities
of peaks are about five times larger than that of the standard CDM simulation
(WFDE), and may correspond to the density of fainter galaxies.

\bigskip
\bigskip
\leftline{\Bold 3. Clustering properties and comparison with observations}
\bigskip
\leftline{\bf 3.1 Power spectrum and evolution}
\smallskip
As an example, Figure 3 shows the evolution of the density
power spectrum $P(k)$ of Model I
in one realization of the L240 simulation.
The choice of this realization is arbitrary, and the features
shown by this example are typical.
We calculate the power spectrum
from the particle distribution at redshift
$z=$ 8, 4, 2, 1, and 0 respectively.
The particle distribution is first converted by the CIC to a density field
on $128^3$ grid points, then
the $P(k)$ is obtained by Fourier transformation of
the density field. Beyond the Nyquist wavenumber, the $P(k)$ is
meaningless because it is seriously affected by the `alias' effect
and the discreteness of particles.
The results are shown in the figure
by different types of symbols for different
redshifts. For comparison, we also plot the power spectrum
predicted by the {\it linear} perturbation theory (solid lines).
The power spectrum calculated
from the particle distribution at $z=8$ is in good
agreement with the input spectrum up to
$k\approx k_N=32 k_0$ (where $k_0$ is the
fundamental wavenumber of the simulation; $k_N$ is the
particle Nyquist wavenumber in the simulation), indicating that
the initial simulation conditions have been properly generated.
At $z\gs 2$,
the density perturbations follow the
linear perturbation theory for $k\ls k_N$.
Thereafter, non-linear
effects become more and more important on scales $\ls k_N$. At
the present epoch ($z=0$), the density perturbations depart from the
linear prediction for $k\gs 6k_0$, showing the importance of
non-linear evolution on these scales.

The power spectrum at the present time can be related to many
observational quantities,
as we shall see below. The evolved power spectrum
obtained from each ensemble of the simulations is presented in
Figure 4, with the
error bars showing the $1\sigma$ scatters between the five realizations.
The open and filled circles show the results
of the power spectra of the mass-density field,
$P_m(k)$, of the L240 and of the
L60 simulations respectively. Below the Nyquist
wavenumber ($k_{N2}= 0.84\impc$) of the L240 simulations,
the spectra of the L240 simulations agree quite well with those of
the L60. Above $k_{N2}$, the power spectra $P_m(k)$ of the L240
simulations are smaller than those of the L60 simulations, which
is partly due to the force softening in the N-body code and
partly due to the regular placement of particles on the grid
points
in generating the initial conditions.
Combining these two sets
of simulations, we can therefore study the clustering properties
of the models on a scale $k\ls k_{N1}=3.3 \impc$ (where $k_{N1}$ is the
Nyquist wavenumber of the L60 simulations).

We have also applied the power spectrum analysis to the spatial
peak distributions.
In the following calculations,
we shall mainly consider the density field
of peaks. The peak-density fields are generated by
the CIC assignment of peaks
to $128^3$ grid points. Their power spectra,
$P_p(k)$, are depicted in Figure 4 by triangles.
The results of the L240 simulations are shown by open symbols, and
those of the L60 by filled ones.
For clarity, the peak power spectra of $P_p(k)$ have
been shifted by a factor of 10
in the figure. Compared with the power spectra
$P_m(k)$ of the mass distributions, the $P_p(k)$ are amplified
by roughly a constant factor $b^2>1$ ($b$ is usually called ``the biasing
factor"). This means that the linear
biasing assumption is approximately
valid in the peak scheme adopted here.
For the peak height thresholds
we specified, the values of $b$ are about
1.2 in Model I, 1.5 in Model II, and
1.6 in Model III. It is interesting to note
that these values are consistent with recent
observational results (Yahil 1988; Kaiser
et al. 1991; Lahav \& Kaiser 1989).

Comparing
the evolved spectra with those predicted by the
linear perturbation theory (the dotted lines in Fig. 4),
one sees that
the effect of the nonlinear evolution
is obvious for $k>0.15\impc$.
We then fit the evolved spectra by simple analytical formulae up to
wavenumber $k_{N1}$, the resolution limit of
the L60 simulations.  The formulae
we use have asymptotically the same functional
form as the linear power
spectra for $k< 0.1\impc$, and are described by
a power-law $k^{-\alpha}$ with  $\alpha \approx 1.3$ for
$k\gs 0.5\impc$. The results of the fit
are shown by the solid lines in Figure 4
(valid for $k\ls k_{N1}$). These fit curves will be used
in the following subsections.

\bigskip
\leftline{\bf 3.2 Two-point correlation functions}
\smallskip
The two-point correlation functions $\xi(r)$ of the simulation particles
and of the peaks are
plotted in Figure 5 for the three models. The
squares show the results for the peaks, and the circles
represent the results for the mass particles. The
functions $\xi(r)$ of the
L60 simulations are shown by the filled symbols, and those
of the L240 by the open ones. In calculating the peak-peak
correlation function, we use the peak number to weigh the
pair counts. Similar to the power spectra (\S3.1),
the correlation functions of the peaks are roughly a factor $b^2$ higher
than those of the underlying mass. The
functions $\xi(r)$ of the L60 and
L240 simulations are in reasonable agreement
on intermediate scales. Because of the
resolution limitation in the simulations, clustering is suppressed
on small scales, as shown by the flattening of $\xi(r)$ on
$r\ls 1\mpc$ in the L240 simulations and on $r\ls 0.3\mpc$
in the L60 simulations. $\xi(r)$ of the L60 simulations
are smaller than those of the L240 simulations on
scales $r\approx 10
\mpc$, which is probably caused by the suppression of
the non-linear clustering
development on scales close to the simulation box size and by the
truncation of the power on scales larger than the size of
the box.

In all three models, the correlation functions have
approximately
a power-law form $r^{-\gamma}$ with $\gamma\approx 1.7$ for
$0.5 \ls r\ls 10\mpc$ (see the dashed line which shows
the power law $r^{-1.8}$). This is true for both the underlying
mass and the peaks. The correlation lengths $r_0$ [defined as the
scale where $\xi(r)$ is 1] of the mass particles are about 4.8, 4.0, and
3.0 $\mpc$ for Model I, Model II and Model III respectively.
The corresponding values for the
peaks
are 6.3, 6.4, and 5.9 $\mpc$.
The results for the peaks, by our design,
are in agreement with the observed correlation
function of galaxies (Davis \& Peebles 1983; de Lapparent et al. 1989).
The data of MESL also suggest that the correlation length
of galaxies is around $6\mpc$ (see below).

The two-point correlation function and the power
spectrum form a Fourier transform pair. Given a
power spectrum, one can easily calculate its correlation function.
The solid and the dotted curves in Figure 5 show $\xi(r)$ for the
peaks and for the underlying mass calculated from the power spectra
shown in
Figure 4 (the solid curves). In these calculations, we have
included the softened part of
the power spectrum $P(k)$ on scales $k\ge k_{N1}$, because
the correlation functions $\xi(r)$
estimated from the pair counts have also been
affected by the same amount
of softening. $\xi(r)$ obtained from $P(k)$ remains
positive up to about $44\mpc$ in Model I, $51\mpc$ in Model II
and $60\mpc$ in Model III.
However, $\xi(r)$ is difficult to detect below 0.01 from direct
pair counting, because of the integral constraint
$\int_0^\infty\xi(r)r^2dr=0$ and the lack of clustering power
outside of the simulation box.

The angular two-point correlation function $\omega(\theta)$,
determined
from large deep surveys of galaxies,
challenges strongly the
standard CDM model (MESL; Picard 1991;
Collins et al. 1992).
Here we
calculate $\omega(\theta)$ for the hybrid models. We convert
the peak correlation functions $\xi(r)$ (the
solid curves in Figure 5)
to $\w$ by using the relativistic
version of the Limber equation (Peebles, 1980 \S 56).
An extrapolation by the power
law $\propto r^{-1.7}$
is made for
$\xi(r)$ on scales $r\ls 0.4\mpc$,
to compensate for the softening effect in the simulations. The
extrapolation is important only on very small
angular scales $\theta\ls 0.07^\circ$. In order to compare the theoretical
$\w$ with the APM observation, the luminosity function and its evolution
model given by MESL are used here. Furthermore, the clustering evolution is
assumed to be stable, i.e., $\xi(r,z)\propto (1+z)^{-3}$,  as in MESL.
In Figure 6, we show
the angular correlation function
$w(\theta)$ of the hybrid models scaled to the Lick
catalog depth (the solid curves).  The APM observational data are
also plotted for comparison. Of the three models,
Model III has the strongest and Model I the weakest power
on large angular scales, as expected because the correlation
functions $\xi(r)$ of peaks are essentially normalized on scales
$r\ls 10\mpc$. Model III is clearly in good agreement with the
$\w$ of the APM survey; Model II is acceptable, considering
that the data for $\w<0.01$  may be not reliable due to
the possible plate-to-plate sensitivity variation in the survey
(MESL). Model I does not have
sufficient power on large scales
to explain the APM data.

\bigskip
\leftline{\bf 3.3 Count-in-cell statistic of galaxies}
\smallskip
Another important observational
result of the large scale structure comes from the count-in-cell
statistic. Assuming that the window function for counting is
$W(\vec r)$, the variance $\sigma^2$ of the counts $N$ in cells
can be easily related to the two-point correlation function
$\xi(r)$:
$$\sigma^2={1\over [\int _{V_\mu}
W(\vec r)d\vec r]^2}\int \xi(|\vec r_1-\vec r_2|)
W(\vec r_1)W(\vec r_2)d\vec r_1 d\vec r_2\,,\eqno(1)$$
or equivalently to the power spectrum $P(k)$:
$$\sigma^2={V_\mu\over (2\pi)^2}\int P(k) |W(\vec k)|^2 d\vec k
\,,\eqno(2)$$
where $W(\vec k)=\int _{V_\mu}
W(\vec r) e^{-i\vec k\cdot \vec r}d\vec r
/[\int W(\vec r)d\vec r]^2$ and $V_\mu$ is a sufficiently large
rectangular volume in which the P(k) is measured.

Efstathiou et al. (1990, hereafter E90) and Saunders et al. (1991)
have applied this statistic to the QDOT redshift survey of IRAS
galaxies. E90 used cubic cells, i.e.,
$$ W(\vec k) ={\sin({l\over 2}k_x)\sin({l\over 2}k_y)
\sin({l\over 2}k_z) \over ({l\over 2}k_x)({l\over 2}k_y)({l\over 2}k_z)}
\,,\eqno(3)$$
where $l$ is the side length of a cell. Saunders et al. (1991) used
Gaussian windows for their measurement. Since the two observations
are closely related and are based on the same database, their constraints
on theoretical models are quite similar. Therefore we test the
hybrid models only against the measurement of E90.

E90 measured the variance $\sigma^2$ using cubic cells of $l=10,\,20,\,30,
\,40,\,{\rm and}\, 60\mpc$. Their results are shown in Figure 7. To see the
predictions of the hybrid models for the variance, we first calculate
$\sigma^2(l)$ analytically by Eqs. (2) and (3) with the power spectra $P(k)$
given by the fits in Figure 4 for galaxies.
The $\sigma^2(l)$ predicted are
shown in Figure 7 by dotted lines. The models have higher $\sigma^2(l)$ at
$l\approx 10\mpc$ and faster decrease of $\sigma^2(l)$ with $l$
than the observation. However E90 measured $\sigma^2$ in redshift space.
Redshift distortion can reduce the clustering
on small scales and enhance that
on large scales (Peebles 1980; Kaiser 1987), thus making $\sigma(l)$
look more flat. Therefore we apply the statistic
$$\sigma^2={\la N^2\ra-\la N\ra^2-\la N\ra \over \la N\ra^2}
\,,\eqno(4)$$
directly to
our simulations, to measure $\sigma(l)$ for the distributions of
peaks in redshift space, where $N$ is the number of galaxies in a
cell. When we transform the particle positions from real space to
redshift space, we simply assume that the `observer' is at
 the center of the simulation
box and is at rest in the comoving frame. We have not attempted
to choose an `observer' having the same environs as ours (e.g.,
the same peculiar velocity relative to the comoving frame, Virgo
cluster being around, etc.), though the statistic may be influenced
by the choice of the observer. The results are given in Figure 7 as the
solid lines.
Indeed, the redshift distortion amplifies $\sigma^2(l)$ by a constant
factor on scales $l\ge 20\mpc$, with an
amplification factor consistent with the
prediction of Kaiser (1987). For $l$
around $10\mpc$, the
scales where the clustering transits
from the nonlinear to the linear regime, the
values of $\sigma^2(l)$
are not affected strongly
by the redshift distortion.
The values
of $\sigma^2(l)$ predicted by the models are higher than the
observational results.
However since the peaks in the
simulations are selected to match the
clustering strength of optical
galaxies, and IRAS galaxies may be less strongly clustered
(e.g., Lahav et al. 1990; Saunders et al. 1992),
the discrepancy can be explained. Since it is not
quite certain how much IRAS galaxies are biased
with respect to
optical galaxies, we
simply shift each solid line in Fig.7  downward by
a constant factor ($b_{OI}^2$)
to fit the observational upper limit of $
\sigma^2(l)$ at $l=10\mpc$. These are the dashed lines in the
figure. The value of $b_{OI}$ is 1.3 for Model I,
1.5 for Model II and 1.4 for Model III. It is interesting to
note that the relative biasing parameter for optical
and IRAS galaxies obtained by Lahav et al. (1990)
is about 1.7.
Compared with the standard CDM model (the open squares),
the hybrid models have much more power
on large scales. Furthermore,
Model II and Model III give better fits to the observation than
Model I.
The results of these three
models are consistent with all of the observed data points
except the one at $l=40\mpc$,
where the models still do not have sufficient
power to reproduce the high $\sigma^2$ value.
Should this observed value
not be a statistical fluke, we would need more excess power on
large scales than the hybrid models have.
Furthermore the power spectrum required would be quite different from
the predictions of the CDM and the hybrid models, because these
models produce smooth curves of $\sigma^2(l)$ without the type
of bump at $l=40\mpc$ seen in the data.
As Fisher et al. (1993) recently pointed out, the value is
significantly in excess of the variance
determined from their 1.2 Jy IRAS redshift
survey.
\bigskip
\leftline{\bf 3.4 The bulk motions on large scales}
\smallskip
Our local universe within a distance of
$\ls 50\mpc$ was found to move coherently
at a velocity of $\sim 600 \kms$ in the direction of
Centaurus
(Lynden-Bell et al. 1988; Faber \& Burstein 1989). This is
one piece of evidence for
the existence of more large-scale
clustering power in the universe than
given by the standard CDM model
(e.g., Vittorio, Juszkiewicz, \& Davis 1986). A recent
reanalysis of the Lynden-Bell et al's data,
using a reconstruction method, gives a bulk motion
velocity of $388\pm 67\kms$
toward $L=177^\circ$, $B=-15^\circ$
for a sphere of radius $40\mpc$ around the Local group
(LG), and a bulk motion of
$327\pm 87\kms$ toward
$L=194^\circ$, $B=5^\circ$,
for a sphere
of radius $60\mpc$ (Bertschinger et al. 1990).

The relation between the bulk motion $V(R)$ and the density power
spectrum $P(k)$ has been extensively discussed in Juszkiewicz et al.
(1990, JVW) and Lahav et al. (1990, LKH). Because the observational
region is centered on the LG, it seems reasonable to study
relevant statistics under the condition that the central parts
in the theoretical models have the same peculiar velocity
as the LG\footnote{\dag} {The condition may still be insufficient
for an `appropriate' observer, since our location in the Universe is
so specific. Several rich clusters and large voids around us may
significantly influence measurement of the bulk motions.}.
The peculiar velocity of the LG is $622\kms$ according to
the COBE observation (Smoot et al. 1991). We model the LG by a
top-hat window of radius $5\mpc$. All information about
bulk motion $\vec V(R)$ is thus contained in the conditional
probability distribution function of $\vec V(R)$, which
depends only on $P(k)$ (JVW; LKH). Using the formulae
derived by JVW and LKH, and using the density power spectra given
in \S 3.1, we show in Figure 8 the predictions of the
hybrid models for the bulk motion. The solid lines are the rms
bulk velocities of the models. For these three models, the
predictions for the bulk velocity are indistinguishable and
all agree well with the statistical
results of Bertschinger et al. (1990).
This result, plus the so-called
`cosmic variance'
which is given by the theoretical upper and lower
limits of the 95\% confidence level (dashed lines),
means that the constraint
on models given by the present observations of
bulk motion is very weak.

\bigskip
\leftline{\bf 3.5 The pairwise velocity dispersion on small scales}
\smallskip
The pairwise velocity dispersion around $1\mpc$ has been considered to
be a strong test of galaxy formation theories. As in the observations,
we calculate the one dimensional velocity
dispersion, i.e., the {\it rms} relative
velocity $\la v_{\|}^2\ra^{1/2}$
along the connection lines of pairs. We
would like to point out that the observational value depends to some
degree on the
model assumed for the pairwise velocity distribution (cf. Peebles 1980).
To have a precise comparison between the model and the observation,
we should use exactly the same procedure to estimate the dispersion
in the simulation as in the observation. Here we ignore the error
possibly caused by the model of the pairwise velocity distribution.
Only the L60 simulations are used for this analysis. In the simulations,
each particle carries less than one peak, that is, one `galaxy'
consists of more than one particle. The pairwise velocity dispersion
of `galaxies' could be very sensitive to the procedures used to
identify them (Couchman \& Carberg 1992; K93).
Since we are unclear which identification procedure is really correct,
here we have measured the velocity  dispersion of the mass and assume
that it is equal to the velocity dispersion of
galaxies (cf. Couchman \& Carlberg 1992).
In Figure 9, we present $\la v_{\|}^2\ra^{1/2}$ for the three models.
Again, error bars are the $1\sigma$ scatters between the five realizations.
$\la v_{\|}^2\ra^{1/2}$ is between 800 and 1100 $\kms$ in Model I,
between 700 and 950 $\kms$ in Model II, and between 600 and 800 $\kms$ in
Model III.
All of these values are significantly higher than
the observational values:
$340\pm 40\kms$ for the CfA survey (Davis \& Peebles 1983);
$250\pm 50\kms$ for the AAT pencil-beam samples (Bean et al. 1983);
$\sim 400\kms$ for the CfA slice survey around the Coma cluster
(de Lapparent, Geller \& Huchra 1988);
and $300\pm 100\kms$ for the KOS pencil-beam survey (Efstathiou \&
Jedredjewski 1984). However there were two statistical studies which
gave a much larger velocity dispersion for galaxies. Efstathiou \&
Jedredjewski (1984) got $540\pm 100\kms$ for the KOSS pencil
beam survey, and Hale-Sutton et al. (1989) obtained $600\pm 140
\kms$ for the AAT sparse pencil-beam survey. The last two values are
in agreement with the predictions of Model II and Model III.

We point out here that almost all N-body
simulations, carried out
to date, give high values
of the pairwise velocity dispersion.
The standard CDM model without bias predicts $\la v_{\|}^2\ra^{1/2}$
$\gs 1000\kms$ (e.g., Davis et al. 1985; Couchman \& Carlberg 1992).
Even the standard CDM model with a high bias $b_g=2\sim 2.5$ (e.g.,
Davis et al. 1985; Park 1991) or a low density ($\Omega=0.2$) CDM model
(Kauffmann \& White 1992),
predicts a $\la v_{\|}^2\ra^{1/2}$
value between $400$ and $600\kms$. As Gelb
et al. (1993) showed recently,
the pairwise velocity dispersion is
very sensitive to the bias parameter; $\la v_{\|}^2\ra^{1/2}\approx 500
\kms$ requires the {\it rms} density contrast $\sigma_8$ in a sphere of
radius $8\mpc$ to be 0.5 in both the flat CDM and the hybrid
(with a contribution of neutrinos
of less than $30\%$ in mass) models.
For the three models studied here,
the values of $\sigma_8$ are
0.86 for Model I, 0.73 for Model II
and 0.63 for Model III. All are larger than $0.5$.
The high values of $\la v_{\|}^2\ra^{1/2}$
in these models are therefore expected.

It may also be possible that
some of the statistical results
for the galaxy pairwise velocity dispersion
listed above represent only a lower limit of the true
value.
As pointed out by Mo, Jing \& B\"orner (1993; hereafter MJB),
the statistical result of the galaxy pairwise velocity
dispersion is very sensitive to sampling effects, such as
the correction of the Virgocentric infall in the CfA sample
and the exclusion of Coma region from the CfA slice.
Without
these corrections, MJB found a value
$450\kms$ for the CfA survey and $\sim 1000\kms$ for
the CfA slice sample. MJB have also analyzed the Southern Sky Redshift
Survey (da Costa et al. 1991) and the 2Jy IRAS galaxy survey
(Strauss et al. 1992).
The pairwise velocity dispersions they found are
$\sim 400\kms$ for the SSRS sample
and $\sim 350\kms$ for the IRAS sample.
{}From these results,
they obtained
$\la v_{\|}^2\ra^{1/2}=500\pm 50\kms$.
Therefore
the discrepancy between the predictions of our
models (especially Models II \& III) and the
observation is alleviated.

The density power spectra in the simulations are normalized by the
COBE quadrupole result.
Taking the lower limit allowed by the
uncertainty ($\sim 25\%$) in the quadrupole
measurement,
one can expect
to find a pairwise velocity dispersion
of $\sim 500\kms$ in Model II and Model III,
in good agreement with
the observation. The modest reduction of the
amplitude of the power spectrum $P(k)$ has,
however, little influence on the results of the previous
sections. As discussed in JMBF, this reduction can
also improve the mass function of rich clusters in Model II
and Model III.

\bigskip
\leftline{\bf 3.6 Filaments, bubbles and voids}
\smallskip
Recent observations have revealed a variety of structures in the universe
(Geller \& Huchra 1988; references therein).
Although the morphology of these
structures has potential importance
for the theories of cosmogony, it is
difficult to
assess the statistical significance
from just a few individual structures
(e.g., voids, filaments, superclusters, bubbles etc.). In this section, we
do not intend to set any constraints on the models from such observations,
because we think only well defined and objective statistics are useful
to constrain models.
Instead, we just show a few simulation
slices of Model II, to give a
qualitative impression on the
frothiness of the hybrid models.

In Figure 10, for example, we show the spatial distributions of
galaxies and
the underlying mass in two square slices of dimension $60\times 60$
$\,(\mpc)^2$ and thickness $18\mpc$ in the L60 simulation. There are
about 900 galaxies and 20,000 mass particles in each slice. Since in the
L60 simulation the peak number carried by each mass particle is
less than 1, the galaxies can be selected by a Monte Carlo experiment,
with the selection probability equal to the peak number of each particle
(see, e.g., Weinberg \& Gunn 1989). The number of galaxies thus
selected is the same as the number of peaks, and the correlation function
of the galaxies was found to be the same as
that given by particle pairs weighted by peak
numbers (\S 3.2).
There are originally $\sim 70,000$ mass particles in each of the
slices. For clarity we have only plotted 30\% of them. We use
thick dots to denote galaxies and use thin points
to denote mass particles. From the plots, it is obvious that the
galaxies delineate the structures of the mass distribution, and
that the distribution of galaxies is more clumpy than that of
the underlying mass (the biasing factor is
$\sim 1.5$, see \S 3.1).
In the left panel
one finds prominent filaments
with lengths $\sim 50\mpc$.
In between the filaments one finds
significant underdensity regions (voids) of
similar sizes. The spatial distribution
shown in the right panel looks much
smoother. Here one finds
many small voids with sizes of
about $10 - 20 \mpc$, surrounded by
filaments or/and walls.

In Figure 11, we show a redshift slice of
the galaxy distribution, which is from one realization
of the
L240 simulation.
For this plot, we assume that the `observer' is sitting
at the center, and the richest cluster in the simulation has
equatorial coordinates
$\alpha=13^h$, $\delta=29^\circ$,
and redshift $cz=8,500 \kms$, the position
of the Coma cluster in the real universe. The slice has
its boundary specified by the declination
range $26.5^\circ\le\delta\le 38.5^\circ$ and by the depth
$cz=12,000 \kms$. The sample of galaxies is constructed by
using a
radial selection function which
corresponds to an apparent-
magnitude limit
$m_B=15.5$.
These selection criteria mimic the CfA slice
survey of de Lapparent, Geller \& Huchra
(1988). However unlike the observation,
the sample does neither have a boundary in right-ascension
nor take into account the galactic extinction effect.
There are pronounced large
empty regions around the rich cluster,
The galaxy distribution in this slice
looks quite similar to that in the CfA
sample, having rich clusters surrounded
by large voids.

\bigskip
\bigskip
\leftline{\Bold 4 Discussion and Conclusions}
\bigskip

The N-body simulations carried out are for the purpose
of testing the models in more detail. They are not
designed to probe the model-parameter space
allowed by present observations. One can perhaps
find many hybrid models compatible
with the observations, by tuning model parameters such
as the cosmological density parameter $\Omega_0$,
the Hubble constant $H_0$, the relative fractions of CDM,
HDM and baryonic components, the shape of the primordial
density spectrum, and the statistic (Gaussian or non-Gaussian)
of the primordial density fluctuations. The models studied
in this paper are favored by an inflationary universe, and by
previous studies based on the linear and non-linear
calculations.

The success of Model II and Model III in matching the
observed large-scale clustering and motions
of galaxies demonstrates that the shape of the initial
density spectrum for structure formation in the universe
is very close to those given by these models.
However, it is very difficult to assess how serious
one should take these specific models.
Model III is not favored by the current
experimental data of neutrinos masses.
However,
the assumptions made on the species and masses of
the dark matter particles in Model II are not in conflict
with currently popular theories of particle physics,
but there is still no compelling evidence
for this model.
But since the
constraints from particle physics are not
stringent, the possibility for the dark matter to have
the properties assumed may not be very small.
Although the two models have very different assumptions
on the properties of the dark matter, their transfer functions
are very similar.
This means that the initial density spectrum given by
these specific models may represent those of a class of
(dark-matter) models, and therefore may have more
validity than the models themselves.

Galaxy formation may be a potential problem for
the hybrid models. Since these models have less power
on small scales than the standard CDM model,
galaxy formation at high redshift is less efficient.
Indeed, our simulation show that significant
nonlinear evolution occurs only after
$z\approx 2$. It is not known if galactic-scale
perturbations can collapse early enough to
account for the existence of quasars and radio galaxies
observed at $z>2$. As discussed by Melott (1990;
see also Buchert \& Blanchard 1993), if quasars
(and radio galaxies) are objects as rare as they
are observed to be, the constraint imposed
by the galaxy-formation time is not very stringent
even in a pure HDM model. To study galaxy formation in the hybrid
model in more detail, it is necessary to have
simulations which incorporate the streaming motion
of neutrinos and the hydrodynamical processes of
baryons on small scales.
Such a simulation
is inevitably much more difficult to carry out
than that for the CDM model. Until now
only few pioneer studies have been done
(e.g. Davis et al. 1992; K93). It is fair to say
that galaxy formation in hybrid models is still
an open question.

Without a proper understanding of the physical processes
of galaxy formation, any biasing procedure for identifying
`galaxies' in simulations is not well justified. The
procedure we used here is based on the biasing
formalism of Bardeen et al. (1986), and is very similar to that
of WFDE for the simulation of the standard CDM model
in large boxes.
For the standard CDM model, it is known that
the `galaxies' selected in this way correspond well
to the real density peaks in high-resolution
simulations (Park 1991). Presumably this result is also true
for the hybrid models, for the structure formation here is
still a bottom-up gravitational hierarchy.
But to justify this needs simulations of high
resolution. However since the biasing factor is
determined by the mass correlation function
(which is, in turn, uniquely determined by normalizing
the initial density spectrum to the COBE
measurement of the MBR fluctuation) and the observed two-point
correlation function of (bright) galaxies, the value
of the biasing parameter obtained for these models
should be reliable, unless the COBE
result is seriously in error. Our results suggest
that a modest bias of galaxy formation is required
for these models to match both the COBE result and the
galaxy clustering on small scales. An interesting question
is whether or not such a bias can be naturally achieved
by the galaxy formation process in these models.

Keeping the above discussed questions in mind,
we now summarize our main conclusions as follows.

1. Hybrid models of the universe which contain
(in mass) $\sim 30\%$ HDM, $\sim 70$\% CDM and
$\gs 1$\% baryons, when normalized to the COBE
quadrupole result of the MBR, provide a reasonable
fit to the observed galaxy clustering and motion
of the universe on large and intermediate scales.

2. For the hybrid models to match both the COBE
results and the observed two-point correlation
function of galaxies, a modest bias, with a biasing parameter
$b_g\sim 1.5$, is needed. This value is in good
agreement with observations.

3. The constraints from different observations
of galaxy clustering and motions are consistent
with each other, supporting that
the large-scale structure of the universe is
from the gravitational instability
of the initial density fluctuations which
have a power spectrum very close to those
predicted by the hybrid models.

4. The most stringent constraint on the models
comes from the angular-correlation function
[$w(\theta)$] of galaxies of the APM
survey (Maddox et al. 1990), and from the
count-in-cell function [$\sigma ^2(l)$]
of IRAS galaxies of the QDOT survey
(e.g. Efstathiou et al. 1990). Of the
three models we studied (see \S2 for definition),
Model III is in good agreement with the
results of the APM survey; Model II is
acceptable, considering the possible
selection effects in the survey;
Model I does not have sufficient power of galaxy clustering on
large scales to match the APM data. The three models
are consistent with the statistical results of the
count-in-cell variance $\sigma^2(l)$,  except for the
observed data of $\sigma^2(l)$ at $l=40\mpc$ which are too high.

5. All of the three models are consistent with the
observations of the bulk motions of galaxies, with
Model III giving the best fit.
The constraint given by this observation is weak.

6. If the velocity bias is negligible,
the pairwise velocity dispersion
$\la v_{\|}^2\ra^{1/2}$ in the models
is uniquely determined
by the normalization to the COBE
quadrupole measurement.
The values of
$\la v_{\|}^2\ra^{1/2}$ are between
800 and 1100$\kms$ in Model I, between 700  and 950$\kms$
in Model II, and between 600 and 800 $\kms$ in Model III. These
values are still larger than the recent statistical result
$\sim 500\kms$
obtained by MJB based on various redshift surveys
of galaxies. However, the
discrepancy is not severe, especially for
Model II and Model III. Using a
modestly reduced (by $\sim 15\%$)
amplitude of the
MBR quadrupole (which is within the COBE detection uncertainty)
to normalize models, one
can pull the model prediction down to $\sim 500\kms$.
This slight change in normalization does not influence
other conclusions of the paper.

7. Voids of the size of Bootes, filaments
of the size of the Great Wall are easily
produced in the hybrid models. The morphology
of the large-scale structure resembles well
that revealed by observations.

8. To have a complete discussion on our simulations, we also
summarize
the results of JMBF on the clustering properties
of rich clusters in the models. They found that
both the coherence length and the correlation
length of the cluster-cluster correlation function in
the hybrid models are larger than in the standard CDM model, and
increase systematically with the power of the initial
density spectrum on large scales. The correlation
length is $r_0=(15.5\pm 0.8)$, $(20.0\pm 0.8)$
and $(23.0\pm 0.9)\mpc$ for Model I, Model II
and Model III, respectively. The corresponding
coherence lengths are about
40, 50, $70\mpc$. Model III matches the
observed correlation function in both
amplitude and coherent length.
Model II is also consistent with the
observations within the theoretical and observational
uncertainties. Model I gives a correlation function
which is too small in both correlation amplitude and
in coherent length.
\bigskip
\bigskip
\leftline {\bf Acknowledgments}
We are grateful to the referee, F.J. Summers, for his critical
comments, to D. Pogosyan and S.P. Xiang for
their discussions on the initial density spectra,
and to R. Valdarnini for discussion on setting
initial conditions in two-component simulations.
S. Maddox is
thanked for kindly sending us the APM $\omega(\theta)$ data
via electronic mail.
L.Z.F. acknowledges
the support of the International Program Development of
the University of Arizona, Y.P.J. acknowledges the supports of
SISSA at the initial stage of the work and
of a World Laboratory Scholarship at the final stage, and H.J.M.
acknowledges the
support
of an SERC Postdoctoral Fellowship.
Y.P.J. and H.J.M. thank the Max-Planck Institut f\"ur
Astrophysik where the work started for
warm hospitality.
\vfill
\eject
\leftline {\Bold References}
\smallskip
\ref
Alimi, J.M., Valls-Gabaud, D., \& Blanchard, A. 1988, A\&A, 206, L11
\ref
Bahcall, N.A., \& Soneira, R.M. 1983, ApJ, 270, 20
\ref
Batuski, D.J., Bahcall, N.A., Olowin, R.P, \& Burns, J.O. 1989, ApJ, 341, 599
\ref
Bardeen, J., Bond, J.R., Kaiser, N., \& Szalay, A.S.
1986, ApJ, 304, 15
\ref
Bean, A.J., Efstathiou, G., Ellis, R.S., Peterson, B.A., \& Shanks, T. 1983,
MNRAS, 205, 605
\ref
Bertschinger, E., Dekel, A., Faber, S.M., Dressler, A.,
\& Burstein, D. 1990, ApJ, 364, 370
\ref
Bond, J.R., Efstathiou, G., Lubin, P.M., \& Meinhold, P.R. 1991,
Phys. Rev. Lett., 66, 2179
\ref
Buchert, T. \& Blanchard, A. 1993, A\&A (in press)
\ref
Collins, C.A., Nichol, R.C., \& Lumsden, L.
        1992, MNRAS, 254, 295
\ref
Couchman, H.M.P.\& Carlberg, R.G. 1992, ApJ, 389, 453
\ref
da Costa, L.N., Pellegrini, P.S., Davis, M., Meiksin, A.,
Sargent, W.L.W., \& Tonry, J.L. 1991, ApJS, 75, 935
\ref
Dalton, G.B., Efstathiou, G., Maddox, S.J., \& Sutherland, W.J.
	1992, ApJ, 390, L1
\ref
Davis, M., Efstathiou, G., Frenk, C. S., \& White, S. D. M. 1985, ApJ, 292, 371
\ref
Davis, M., \& Efstathiou, G. 1988, In  Large-Scale Motions in the
Universe, eds. V.C. Rubin and G.V. Coyne
(Princeton: Princeton Univ. Press)
\ref
Davis, M., \& Peebles, P.J.E. 1983, ApJ, 267, 465
\ref
Davis, M., Summers, F.J., \& Schlegel, D. 1992, Nat, 359, 393
\ref
de Lapparent, V., Geller, M.J., \& Huchra, J.P. 1988, ApJ, 332, 44
\ref
Dekel, A. \& Rees, M.J. 1987, Nat, 326, 455
\ref
Efstathiou, G., Bond, J.R., \& White, S.D.M.
1992, MNRAS, 258, 1p
\ref
Efstathiou, G., Davis, M., Frenk, C.S. \& White, S.D.M. 1985,
ApJS, 57, 241
\ref
Efstathiou, G., \& Jedrzejewski, R.I. 1984, Adv. Space Res., 3, 379
\ref
Efstathiou, G., Kaiser, N., Saunders, W., Rowan-Robinson, M.,
Lawrence, A., Ellis, R.S.,
\& Frenk, C.S. 1990, MNRAS, 247, 10p
\ref
Faber, S.M. \& Burstein, D. 1988, in Large-Scale Motions
in the Universe, eds. V.C. Rubin \& G.V. Coyne
(Princeton: Princeton Univ. Press)
\ref
Fang, L.Z., Xiang, S.P. \& Li, S.X. 1984, A\&A, 140, 77
\ref
Fisher, K.B., Davis, M., Strauss, M.A., Yahil, A., \& Huchra, J.P. 1993
	ApJ, 402, 42
\ref
Gelb, J.M., Gradwohl, B.-A., Frieman, J.A. 1993, ApJ, 403, L5
\ref
Geller, M.J., \& Huchra, J. 1988, in Large-Scale Motions
in the Universe, eds. V.C. Rubin \& G.V. Coyne
(Princeton: Princeton Univ. Press)
\ref
Groth, E.J., \& Peebles, P.J.E. 1977, ApJ, 217, 385
\ref
Hale-Sutton, D., Fong, R., Metcalfe, N., Shanks, T. 1989, MNRAS,
237, 569
\ref
Hockney, R.W.  \& Eastwood, J.W. 1981, Computer simulations using particles.
Mc Graw-Hill
\ref
Holtzman, J.A. 1989, ApJS, 71, 1
\ref
Huchra, J.P., Henry, J.P., Postman, M., \& Geller, M.J. 1990, ApJ, 365, 66
\ref
Jing, Y.P., Mo, H.J., B\"orner, G., \& Fang, L.Z. 1993,
ApJ, 411, 450 (JMBF)
\ref
Jing, Y.P., \& Valdarnini, R. 1993, ApJ, 406, 6
\ref
Juszkiewicz, R., Vittorio, N., \& Wyse, R. 1990, ApJ, 349, 408 (JVW)
\ref
Kaiser, N. 1984, ApJ, 284, L9
\ref
Kaiser, N. 1987, MNRAS, 227, 1
\ref
Kaiser, N., Efstathiou, G., Ellis, R., Frenk, C., Lawrence, A.,
Rowan-Robinson, M. 1991, MNRAS, 252, 1
\ref
Kaiser, N., \& Lahav, O. 1989, MNRAS, 237, 129
\ref
Kauffmann G., White S. D. M., 1992, MNRAS, 258, 511
\ref
Klypin, A., Holtzman, J., Primack, J. \& Reg\"os, E. 1993, ApJ, (submitted)
\ref
Klypin, A.A., \& Kopylov, A.I. 1983, SvA, 9, L41
\ref
Lahav, O., Kaiser, N., \& Hoffman, Y. 1990, ApJ, 352, 448 (LKH)
\ref
Lahav, O., Nemiroff, R.J., and Piran, T. 1990
ApJ, 350, 119
\ref
Little, B., Weinberg, D.H., \& Park, C. 1991, MNRAS, 253, 295
\ref
Loveday, J., Efstathiou, G., Peterson, B.A., \& Maddox, S.J.
1992, ApJ, 390, 338
\ref
Lynden-Bell, D., Faber, S.M., Burstein, D., Davies, R.L.,
Dressler, A., Terlevich, R.J., \& Wegner, G. 1988,
ApJ, 326, 19
\ref
Maddox, S.J., Efstathiou, G., Sutherland, W.J., \& Loveday, J.
1990, MNRAS, 242, 43p
\ref
Melott, A.L. 1986, Phys. Rev. Lett., 56, 1992
\ref
Melott, A.L. 1990, in AIP Conf. Proc. 222, After the First Three
Minutes, ed. S.S. Holt, C.L. Bennett, \& Trimble, V. (New York: AIP)
\ref
Mo, H.J., Jing, Y.P., \& B\"orner, G. 1993, MNRAS, (in press) (MJB)
\ref
Mo, H.J. \& Lahav, O. 1993, MNRAS, 261, 895
\ref
Mo, H.J., Peacock, J.A. \& Xia, X.Y. 1993, MNRAS, 260, 121
\ref
Nichol, R.C., Collins, C.A., Guzzo, L., \& Lumsden, S.L.
	1992, MNRAS, 255, 21p
\ref
Olive, K.A., Schramm, D.N., Steigman, G. \& Walker, T.P.
1990, Phys. Lett., B236, 454
\ref
Park, C. 1991, MNRAS, 251, 167
\ref
Peebles, P.J.E. 1980, Large-Scale Structure of the Universe,
(Princeton: Princeton Univ. Press)
\ref
Picard, A. 1991, ApJ, 368, L7
\ref
Postman, M., Geller, M.J., \& Huchra, J.P. 1986, AJ, 91, 1267
\ref
Postman, M., Huchra, P., \& Geller, M. 1992, ApJ, 384, 404
\ref
Primack, J.A. 1992, preprint-SCIPP 92/51
\ref
Saunders, W., Frenk, C.S., Rowan-Robinson M., Efstathiou, G.,
Lawrence, A., Kaiser, N., Ellis, R., Crawford, J., Xia X-Y.,
and Parry, I. 1991, Nat, 349, 32
\ref
Saunders, W., Rowan-Robinson, M., \& Lawrence, A. 1992, MNRAS, 258, 134
\ref
Shafi, Q. \& Stecker, F.W. 1984, Phys. Rev. Lett., 53, 1292
\ref
Smoot, G.F., et al. 1991, ApJ, 371, L1
\ref
Smoot, G.F., et al. 1992, ApJ, 396, L1
\ref
Strauss, M.A., Huchra, J.P., Davis, M., Yahil, A., K.B. Fisher, \&
Tonry, J. 1992, ApJS, 83, 29
\ref
Taylor, A.N. \& Rowan-Robinson, M. 1992, Nat, 359, 396
\ref
Valdarnini, R. \& Bonometto, S.A. 1985, A\&A, 146, 235
\ref
Walker, T.P., Steigman, G., Schramm, D.N.,
Olive, K.A. \& Kang, H.S. 1991, ApJ, 376, 51
\ref
van Dalen, A., \& Schaefer, R.K. 1992, ApJ, 398, 33
\ref
Vittorio, N., Juszkiewicz, R., \& Davis, M. 1986, Nat, 323, 132
\ref
Weinberg, D.H. \& Gunn, J.E. 1990, MNRAS, 247, 260
\ref
White, S.D.M., Frenk, C.S.,
Davis, M., \& Efstathiou, G. 1987, ApJ, 313, 505
(WFDE)
\ref
Wright, S. et al. 1992, ApJ, 396, L13
\ref
Yahil, A. 1988, In Large-Scale Motions in the
Universe, eds. V.C. Rubin and G.V. Coyne (Princeton: Princeton Univ. Press)
\ref
Xiang, S. \& Kiang, T. 1992, MNRAS, 259, 761
\vfill
\eject
\leftline{\Bold Figure captions}
\smallskip
{\bf Fig.1.} The density power spectra of the hybrid models at
redshift $z=8$, normalized by the COBE quadrupole result of the
microwave background radiation.

{\bf Fig.2.} The influence of the free-streaming motion on
the power spectrum evolution.
(a) The power spectra $P(k)$ estimated from the
particle distributions of the NS simulation (long-dash curves) and the
S simulation (dotted curves) at $z=8$, 4, 2, 1, and 0 (from the bottom to
the top); (b) The power spectra $P(k)$ of cold
(solid curves)
and hot particles (dashed curves) in the
two-component simulation  at $z=8$, 4, 2, 1, and
0 (from the bottom to the top).
The power spectra of the NS simulation are plotted (long-dash
curves) for comparison.
The wavenumber is in units of the fundamental wavenumber
$k_0=2\pi/60\impc$.

{\bf Fig.3.} The evolution of density power spectrum $P(k)$
in one realization of the
L240 simulation of Model I. Different symbols show the power
spectra at redshift $z=8$, 4, 2, 1, and 0 [$P(k)$ is a decreasing
function of $z$]. The solid lines are the predictions for
{\it linear} perturbation evolution. $k$ is in units of
the fundamental wavenumber $k_0=2\pi/240\impc$.

{\bf Fig.4.} The evolved power spectra of
peaks (triangles) and
underlying mass (circles) at redshift $z=0$. For clarity, the
spectra of the peaks are shifted vertically by a factor of 10.
The open and filled symbols show the results of the L240 and of
the L60 simulations respectively. The dotted lines are the
{\it linear} predictions for the density power spectra. The
solid lines are our fits of the evolved spectra. The fit results
are sufficiently accurate for $k_{N1}< 3.3\mpc$, the resolution
limit of the L60 simulation.

{\bf Fig.5.} The spatial two-point correlation functions of
mass (circles) and of peaks (squares).
The open and filled symbols show the results of the L240 and of
the L60 simulations respectively. The solid and dotted curves are the
Fourier transforms of the fitted spectra.

{\bf Fig.6.} The angular two-point correlation functions of galaxies
in the hybrid models, scaled to the Lick catalogue depth, are compared
with the observational data of the APM survey (Maddox et al. 1990;
the APM data were kindly provided by S. Maddox).

{\bf Fig.7.} The count-in-cell variances $\sigma^2(l)$ in cubic
volumes of sides $l$. The observational data, from Efstathiou
et al. (1990), are plotted as squares, together with error bars
representing the 95\% confidence level. The dotted curves show the
variances of peaks in {\it real} space; the
solid curves show the $\sigma^2(l)$ of peaks in {\it redshift} space;
the dashed curves are the solid ones shifted down
so that the $\sigma^2(l)$
equals the observational upper limit at $l=10\mpc$. The open squares
are the prediction of the standard CDM model.

{\bf Fig.8.} The bulk motion of a sphere as a function of its radius
$R$. The solid lines show the theoretical {\it rms} values, and the
dashed ones show the upper or lower limits at the
95\% confidence level.
The filled squares and their error bars are from the statistical
analysis of Bertschinger et al. (1990).

{\bf Fig.9.} The pairwise velocity dispersion along the connection
lines of pairs of particles, as a function of their separation.

{\bf Fig.10.} Two examples of spatial galaxy ({thick dots}) and
underlying mass ({thin points}) distributions in Model II. The
distributions are projected on a rectangular slice of $60\times
60\,(\mpc)^2$. The thickness of the slice is $18\mpc$.

{\bf Fig.11.} The galaxy distribution in a redshift slice of
$26.5^\circ\le\delta\le 38.5^\circ$ in declination,
constructed from one realization of the
L240 simulation of Model II.
The slice contains the richest cluster of the simulation, which
is set to be at the position of Coma cluster in the universe.
The distribution
is subjected to the radial selection function of the CfA slice
survey. The outer boundary of the slice is 12,000$\kms$.

\vfill
\eject
\end